\documentclass[graybox]{svmult}

\usepackage{mathptmx}       
\usepackage{amsmath,amssymb,amsfonts}
\usepackage{bm,graphicx,color}
\usepackage{helvet}         
\usepackage{courier}        
\usepackage{type1cm}        
\usepackage{makeidx}         
\usepackage{graphicx}        
\usepackage{multicol}        
\usepackage[bottom]{footmisc}

\newcommand{\beq}{\begin{equation}}
\newcommand{\eeq}{\end{equation}}
\newcommand{\beqa}{\begin{eqnarray}}
\newcommand{\eeqa}{\end{eqnarray}}
\newcommand{\BK}{\bm{k}}

\makeindex             

\begin{document}

\title*{Material-Specific Investigations of Correlated Electron Systems}
\author{Arno P.\ Kampf, Marcus Kollar, Jan Kune\v{s}, Michael Sentef, and Dieter Vollhardt}
\institute{Arno P.\ Kampf \at Theoretical Physics III, Center for Electronic
          Correlations and Magnetism, Institute of Physics, 
          University of Augsburg, D-86135 Augsburg, Germany, \email{arno.kampf@physik.uni-augsburg.de}}
%
%
\maketitle

\abstract{
We present the results of numerical studies for selected materials with
strongly correlated electrons using a combination of the local-density
approximation and dynamical mean-field theory (DMFT). For the solution of the DMFT equations a continuous-time quantum Monte-Carlo algorithm was employed. All simulations were performed on the supercomputer HLRB II at the Leibniz Rechenzentrum in Munich. Specifically we have analyzed the pressure induced metal-insulator transitions in Fe$_2$O$_3$ and NiS$_2$, the charge susceptibility of the fluctuating-valence elemental metal Yb, and the spectral properties of a covalent band-insulator model which includes local electronic correlations.
}

\section{Introduction}
\label{intro}
The basic concepts of solid-state physics explain the physical properties of
numerous materials such as simple metals, semiconductors,
and insulators. For materials with open $d$ and $f$ shells, where
electrons occupy narrow orbitals, the additional understanding of the role of electron-electron interactions is crucial. In transition metals such as vanadium or iron, and also in their
their oxides, electrons experience a strong Coulomb
repulsion because of the spatial confinement in their respective orbitals.
Such strongly interacting or ``correlated'' electrons cannot be
described as embedded in a static mean field generated by the
other electrons \cite{pt,Imada98,Georges96}.
The $d$ and $f$ electrons have internal degrees of freedom (spin,
charge, orbital moment) whose interplay leads to many remarkable
ordering phenomena at low temperatures. As a
consequence of the competition between different ordering phenomena, strongly correlated electron systems are very
sensitive to small changes in their control parameters
(temperature, pressure, doping, etc.), resulting in strongly
nonlinear responses, and tendencies to phase separate or to form
complex patterns in chemically inhomogeneous situations. 

Understanding the metal-insulator transition (MIT) in strongly
correlated electron systems has been one of the
key topics in condensed matter theory since the 1930's.
It was realized early on that the local Coulomb repulsion plays
a decisive role in the physics of an important class of materials
known as Mott insulators, examples of which are many transition-metal
oxides.  While the role of correlations due to the on-site Coulomb repulsion
has been appreciated for over half a century only quite recently the theory
progressed to the point where they can be treated quantitatively.

In the last two decades, a new approach to electronic lattice models,
the dynamical mean-field theory (DMFT), has led to new analytical and
numerical techniques to study correlated electronic systems \cite%
{pt,Georges96}. This theory -- initiated by Metzner
and Vollhardt in 1989 -- is exact in the limit of infinite
dimensions ($d=\infty $) \cite{Metzner89a}. In this limit, the lattice
problem  reduces to a single-impurity Anderson model with a self-consistency condition \cite{Georges92a,Jarrell92}.
After the initial studies of conceptually simple models,
DMFT has advanced to models for real materials
and only quite recently to purely numerical Hamiltonians obtained from
bandstructure calculations, an approach which goes by the name LDA+DMFT.

\section{Computational Method}
\label{comp}
\subsection{The LDA+DMFT Approach}

In the LDA+DMFT approach \cite{ldadmft} the LDA band structure,
represented by a one-particle Hamiltonian $H_{\mathrm{LDA}}^{0}$, is
supplemented with the local Coulomb repulsion, while the correction
for double-counting the effect of the local interaction is absorbed 
in $\hat{H}_{\mathrm{LDA}}^{0}$,  
\begin{eqnarray}
\label{eq:hubbard}
\hat{H} &=&\hat{H}_{\mathrm{LDA}}^{0}+\sum_{m,m'}\sum_{i}U_{mm'}\hat{n}%
_{im}\hat{n}_{im'}.
\end{eqnarray}%
Here, $i$ denotes a lattice site; $m$ and $m'$ enumerate
different orbitals on the same lattice site. In the present implementation
we use an approximate form of the local interaction consisting
of products of the occupation operators $\hat{n}_{im}$. 

During the last ten years, DMFT has proved to be a successful approach
for investigating Hamiltonians with local interactions as in Eq.\ (\ref{eq:hubbard})\cite{Georges96}. DMFT treats the local dynamics exactly while neglecting
the non-local correlations. In this non-perturbative
approach the lattice problem is mapped onto an effective single-site
problem which has to be solved self-consistently together with the
momentum-integrated Dyson equation connecting the self energy
$\bm{\Sigma}$ and the Green function $G$ at frequency $\omega$: \vspace{-0.5cm}
\begin{eqnarray} 
G_{mm^{\prime }}(\omega
)=\!\frac{1}{V_B}\int{{\ d^{3}}{k}}
\!\!
\!& \left[ \;(\omega +\mu)
 {\bf 1}-H_{\mathrm{LDA}}^{0}
(\bm{k}%
) -\bm{\Sigma}(\omega )\right]^{-1}_{ mm^{\prime }} .\;& \label{Dyson}
\end{eqnarray}
Here, ${\bf 1}$ is the unit matrix, $\mu$ the chemical potential, and
$\bm{\Sigma}(\omega)$ denotes the self-energy matrix which is nonzero only
between the interacting orbitals. $[...]^{-1}$ implies a matrix inversion
in the space with orbital indices $m, m'$.
The integration extends over the
Brillouin zone with volume $V_{B}$.

\subsection {QMC Method}
The single-site impurity problem can be fomulated without explicit construction 
of the fermionic bath using the effective action formalism 
\cite{Georges92a,Jarrell92}. The effect of the bath is represented by the
frequency-dependent hybridization function  $\Delta(\omega)$ given implicitly by
\begin{eqnarray}
G^{-1}_m(\omega )=\omega+\mu-\epsilon_m-\Delta_m(\omega)-\Sigma_m(\omega),
\end{eqnarray}
leading to the action
\begin{eqnarray}
S[\psi^{\ast},\psi^{\phantom\ast},\Delta]
=\int_0^{\beta} d\tau \sum_{m}\psi_m^{\ast}(\tau)\left(\frac{\partial}{\partial \tau}-\mu+\epsilon_m\right)\psi_m(\tau) 
\nonumber\\
+\sum_{m,m'}U_{mm'}\psi _{m}^{\ast }(\tau )\psi_{m}^{\phantom\ast }(\tau )\psi_{m^{\prime}}^{\ast}(\tau)\psi_{m'}^{\phantom\ast}(\tau) \nonumber\\
+\int_0^{\beta}d\tau \int_0^{\beta} d\tau'\sum_m\psi_m^{\ast}(\tau)\Delta_m(\tau-\tau')\psi_m^{\phantom\ast}(\tau'),
\end{eqnarray}
and the local Green function is obtained by a functional integral over the Grassmann variables $\psi^{\phantom\ast}$ and $\psi^{\ast}$
\begin{eqnarray} 
\label{eq:siam}
G_{m}(\tau-\tau')=-\frac{1}{\mathcal{Z}}\int \mathcal{D}[\psi ]\mathcal{D}%
[\psi ^{\ast }]\psi _{m}^{\phantom\ast }(\tau)\psi _{
m}^{\ast }(\tau')e^{S[\psi ,\psi ^{\ast
},\Delta]}. \label{siam}
\end{eqnarray}%
Here, $
\mathcal{Z}=
\int \mathcal{D}[\psi ]
\mathcal{D}[\psi ^{\ast}]
\exp(S[\psi ,\psi ^{\ast },\Delta])
$ 
is the
partition function. 
Such a fomulation in terms of an effective action in imaginary time is well suited for quantum Monte-Carlo (QMC) methods. In the present work
we have used predominantly the continuous-time hybridization expansion QMC algorithm \cite{Werner06}, which
consists in expanding the exponential in (\ref{eq:siam}) in powers of the hybridization function $\Delta$
and sampling the contributions using a QMC random walk,
on which almost all of the required CPU time is spent upon during the iterative solution of the DMFT equations. 

The computational effort of the QMC algorithm based on the expansion of the effective impurity action in the impurity-bath hybridization scales as the cube of the matrix size \cite{Gull07}, which is determined by the mean value of the order of the hybridization expansion. The required order of the expansion scales linearly with inverse temperature, and even decreases upon an increase in the electron-electron interaction strength and is thus an excellent method for the regime of strong correlations. Concerning the cluster extension of DMFT, however, the effort of the full matrix code grows exponentially with the cluster size. Therefore the hybridization expansion solver is efficient for small clusters, but calculations for clusters larger than four sites are prohibitively costly.

From a numerical perspective the continuous-time QMC is perfectly suited for runs on the HLRB II. It is almost parallel by definition, and delayed update optimizations \cite{Alvarez08} are irrelevant to the code since rank-one updates, i.e. single spin flips, are not required. Most of the CPU time is spent on the calculation of a trace, for which the multiplication of small matrices is needed. These matrices fit in the cache and thus the code scales basically linearly with the number of CPUs used. 

The linear scaling in theory is slightly affected in practice by a small overhead due to the thermalization sweeps, which need to be performed on each CPU separately, and more importantly by the data collection procedure of the ALPS tools \cite{alps}, if more than around 300 CPUs are involved. Our typical jobs, however, are rather small scale jobs with 32 to 256 CPUs and CPU times of 10 to 100 CPU hours per DMFT iteration, with a typical number of 20 DMFT iterations. The performance measured for multiorbital single-site calculations is around 100 to 200 MFlops/core, while it fluctuates between 100 and 2000 MFlops/core in simulations of small clusters using the full matrix code. In a typical scaling test for a 2-site cluster DMFT the total CPU time varies between 27 CPU-h on 32 CPUs and 31 CPU-h on 128 CPUs. At the same time the performance slightly decreases from 128 MFlops/core on 32 CPUs to 120 MFlops/core on 128 CPUs. Thus for a typical job, the run time and the performance per core scale almost linearly with the number of CPUs, with a slight decrease of around ten percent upon increasing the number of CPUs by a factor of four.

\begin{figure}[b]
\begin{center}
\includegraphics[scale=0.45,angle=270,clip]{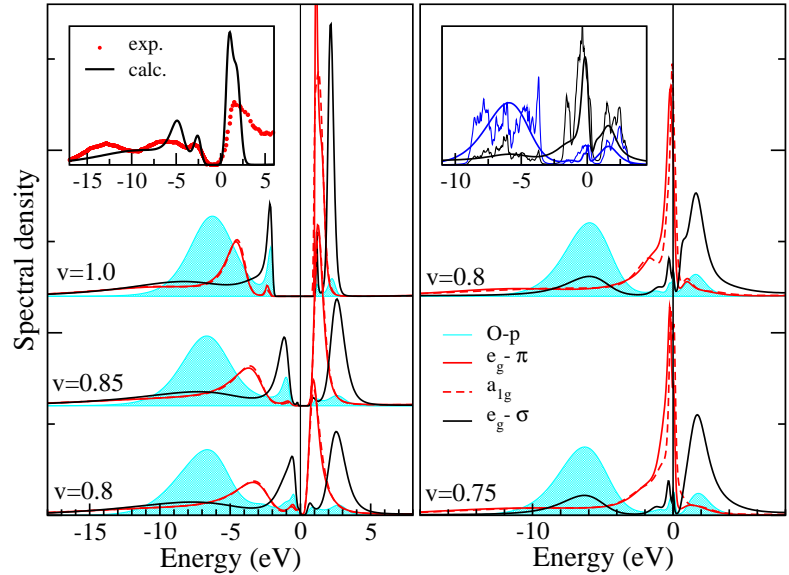}
\end{center}
\caption{Pressure evolution of the paramagnetic single-particle spectra 
(T=580~K). The high-spin solutions are shown in the left and the low-spin solutions
in the right panel. All curves are normalized to unity, and $v$ denotes the volume relative to the volume at ambient pressure.
Left inset: comparison of experimental PES \cite{lad-89}
and IPS \cite{ciccacci-91} data (points) to the $d$-spectra
(solid line) with 0.6 eV Gaussian broadening which is comparable to the experimental resolution.
Right inset: comparison of the v=0.8 DMFT (smooth curves) and
LDA spectra (jagged curves): total $d$-electron spectra (black), the total oxygen $p$-electron spectra
(blue). Figure taken from Ref.\ \cite{fe2o3}.}
\label{fig:1}       
\end{figure}
The computational requirements of the continuous-time QMC are rather simple. It basically needs fast processors in parallel. Not a lot of memory is needed for the QMC and, since the random walks on different CPUs are independent of each other, fast network communications are not required either.

\section{Results and Discussion}
\label{res}
\subsection{Pressure-Driven Metal-Insulator Transition in Hematite}
\label{hem}
An important example for of the MIT
is the pressure driven transition seen in MnO \cite{yoo}, BiFeO$_3$ \cite{gav-08} or Fe$_2$O$_3$ \cite{fe2o3}, which is accompanied by a change of the local spin state (high spin (HS) to low spin (LS) transition).
Understanding the pressure-driven HS-LS transition
and its relationship to
the MIT and structural and/or volume changes
is in fact relevant to a broader class of oxides, often with
geophysical implications.

We have studied the spin transition and the MIT in
hematite ($\alpha$-Fe$_2$O$_3$) under pressure
using the LDA+DMFT approach \cite{ldadmft} including the
effects of temperature and magnetic long-range order (LRO). 
At ambient conditions, hematite has the corundum structure and is an antiferromagnetic (AFM) insulator with a N\'eel temperature $T_N$=956~K \cite {Sgull-51}. The iron ions have a formal
Fe$^{3+}$ valence with five $d$ electrons giving rise to a local HS state.
Photoemission spectroscopy (PES) classified hematite as a charge-transfer
insulator \cite{Fujimori-86,lad-89}. A charge gap of 2.0-2.7~eV was inferred from $dc$ conductivity data \cite{mochizuki}.
Under pressure, a first-order phase transition is observed at approximately 50~GPa
at which the specific volume
decreases by almost 10\% and the crystal
symmetry is reduced (to
the Rh$_2$O$_3$-II structure) \cite {Pasternak-99,Rozenberg-02,Liu-03}.
The high-pressure phase is characterized by metallic conductivity
and the absence of both magnetic LRO and
the HS local moment \cite {Pasternak-99}. Badro {\it et al.}
showed that the structural transition actually precedes the electronic transition,
which is, nevertheless, accompanied by a sizable reduction of the bond lengths \cite{Badro-02}.
\begin{figure}[b]
\begin{center}
\includegraphics[scale=.45,angle=270,clip]{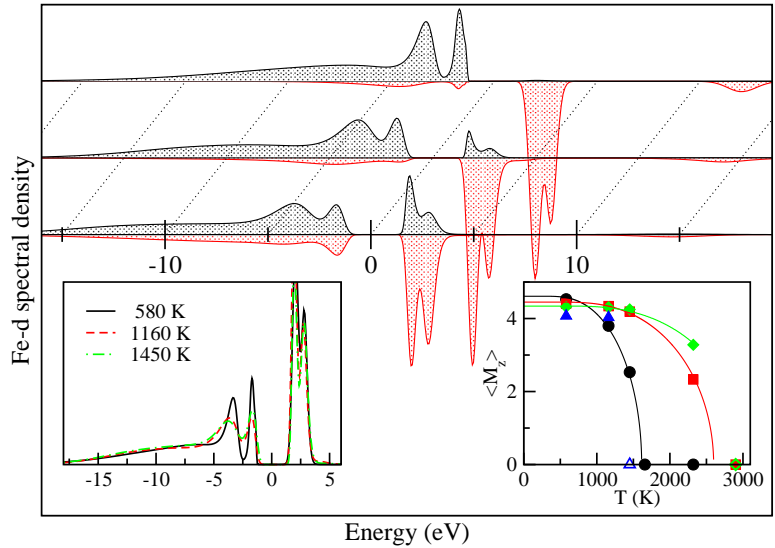}
\end{center}
\caption{Spin-polarized Fe-$d$ spectra at
ambient pressure for 580~K, 1160~K, and 1450~K (top to bottom). Left
inset: the same spectra averaged over spin. Right inset:
staggered magnetization vs.\ temperature for v=1.0 (circles; black),
v=0.9 (squares; red), v=0.85 (diamonds; green), and v=0.8 (triangles; blue), where $v$ denotes the volume relative to the volume at ambient pressure.
For empty symbols only the low-spin solution exists. The lines represent mean-field fits.  Figure taken from Ref.\ \cite{fe2o3}.}
\label{fig:2}
\end{figure} 

The calculations started with the construction of Wannier based Hamiltonians from non-magnetic
LDA bandstructures for various specfic volumes spanning the range of experimentally accessible pressures.
Given the nominal $d^5$ configuration of the Fe-$d$ shell in hematite, a transition between
the high-spin $S=5/2$ state and the low-spin $S=1/2$ state is expected. Our LDA+DMFT calculations were first perfomed
in the paramagnetic phase and exhibited indeed such a HS-LS transition observed as a discontinuous drop
of the expectation value $\langle S_z^2 \rangle$. The HS-LS transition is accompanied by the disappearance of 
the charge gap and a substantial change in the single-particle spectrum (Fig. \ref{fig:1}).
The comparison to iso-electronic MnO, which exhibits a similar HS-LS/insulator-metal transition, reveals
a striking difference between the two materials. While the transition in MnO sets on before the charge gap is closed
due to the pressure-induced increase in bandwidth and crystal-field splitting, the transition in hematite is instead characterized
by a continuous closing of the gap followed by an abrupt spin transition. Comparing these two materials we have
identified the microscopic mechanism behind the observed transitions \cite{fe2o3}.

A further issue addressed the question of long-range magnetic order. The occurence of magnetic order in strongly correlated 
oxides is important also in the broader context of first-principles electronic structure methods such as LDA. In several
such materials a gap appears in the anti-ferromagnetic LDA solutions, while the non-magnetic solutions are metallic.
This is sometimes interpreted as antiferromagnetism causing the opening of the gap.  
Our results explicitly showed that magnetic order has in fact only a marginal effect on the single-particle
spectrum and the size of the charge gap does not change between the paramagnetic and the antiferromagnetic phase (see Fig. \ref{fig:2}).
We also clarified that the presence of antiferromagentic order has only a marginal effect on the HS-LS/insulator-metal transition \cite{fe2o3}.

\subsection{Fluctuating Valence and Valence Transition of Yb under Pressure}
\begin{figure}[t]
\begin{center}
\includegraphics[scale=.38,angle=270]{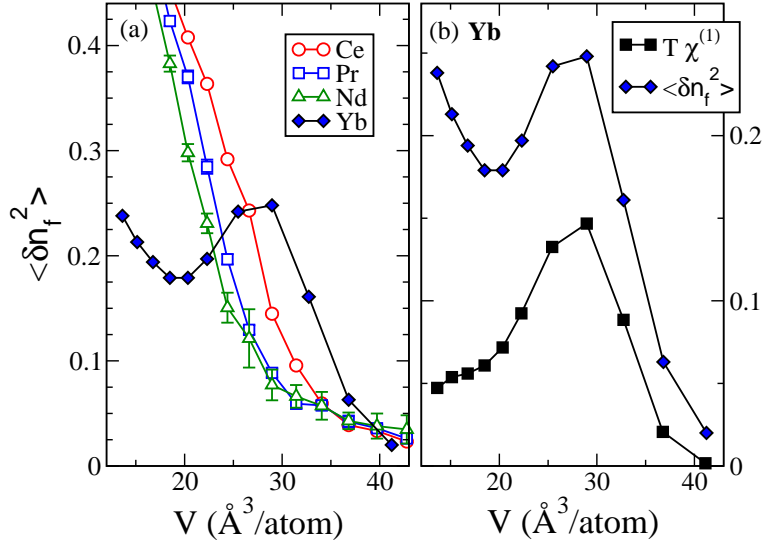}
\end{center}
\caption{(a) Charge fluctuations quantified by the mean-square deviation of the $f$ occuancy in Yb, Ce, Pr, and Nd
at 630 K. If not shown error bars are smaller than the symbol size.
(b) Equal-time charge fluctuations (diamonds, the same as in left panel) compared
to their imaginary-time average given by the product of temperature and the static susceptibility $T\chi^{(1)}$ (squares). Figure taken from Ref.\ \cite{yb}.}
\label{fig:3}
\end{figure}
The valence state of rare-earth atoms in lanthanide compounds also affects their physical properties. The determination of the lanthanide valence
from first principles and the description of the 4$f$ electrons
have been long-standing challenges due to the duality between their atomic character, with a strong local electron-electron interaction, and their itinerant
character due to the lattice periodicity.
Theories based on a two-species picture, which treat part
of the 4$f$ electrons as atomic and the rest as itinerant, succeeded
in reproducing the trends across the lanthanide series for compounds with
integer valence \cite{strange_nature}. Nevertheless, besides being conceptually unsatisfactory,
the two-species picture cannot describe transitions between different valence states
as well as the heavy-fermion behavior of the charge carriers.

We have studied the valence transition in the elemental Yb metal under applied pressure.
Yb and Eu in their elemental form behave quite differently from the other lanthanides. 
If we define the valence as the number of electrons participating in bonding,
the majority of the lanthanide series is trivalent, however for Yb and Eu the $3+$ and $2+$ valence
states are close to degeneracy with the $2+$ state being more stable at ambient conditions \cite{strange_nature}.
This results in a number of anomalous properties, such as a larger molar volume compared to the general trend 
in the lanthanide series, and a lower bulk modulus \cite{Takemura85}.
The thermal-expansion coefficient of Yb is three times larger
than for most other lanthanides \cite{Barson57}.

In accord with the experimental data our calculations reveal a continuous decrease of 
the $f$-shell occupancy, which corresponds to a crossover from the $f^{14}$ to the $f^{13}$ local charge state as was verified by calculating the single-particle spectra \cite{yb}. The change of the $f$ occupancy
is connected to a transfer of electrons to $spd$-bands. The interesting question arises what the 
observable differences are between such self-doping in Yb and the 
rare-earth materials without valence-state degeneracy. The answer is shown in Fig.~\ref{fig:3}
where the mean-square deviations of the $f$-occupancy and the static local charge susceptibilites as a function of 
specific volume are compared for Ce, Nd, Pr, and Yb. The former three exhibit a monotonous increase
of the charge fluctuations with pressure and a small charge susceptibility.
The origin of the increasing charge fluctuations is the growing frequency of hopping processes involving the $f$ electrons.
However, these processes are energetically costly due to the on-site Coulomb repulsion and thus very short lived, 
which leads to the small charge susceptibility. The behavior of Yb is strikingly different.
Besides the increase of charge fluctuations at the smallest volumes, which has essentially the same origin as in the other materials,
the charge susceptibility vs.\ volume exhibits a local maximum approximately in the middle of the valence-transition region \cite{yb}. This feature is directly related to the quasi-degeneracy
of the $f^{13}$ and the $f^{14}$ states. As a consequence, the charge fluctuations are energetically
favorable and thus lead to a large susceptibility.
The observation of the large charge susceptibility provides an explanation for the reported softness of Yb in the valence transition
region.

We conclude that the observed behavior is a common and distiguishing feature of fluctuating
valence systems characterized by the quasi-degeneracy of local charge states.

\subsection{Metal-Insulator Transition in NiS$_{\bm{2-x}}$Se$_{\bm{x}}$}
\begin{figure}
\sidecaption
\includegraphics[scale=.45,clip]{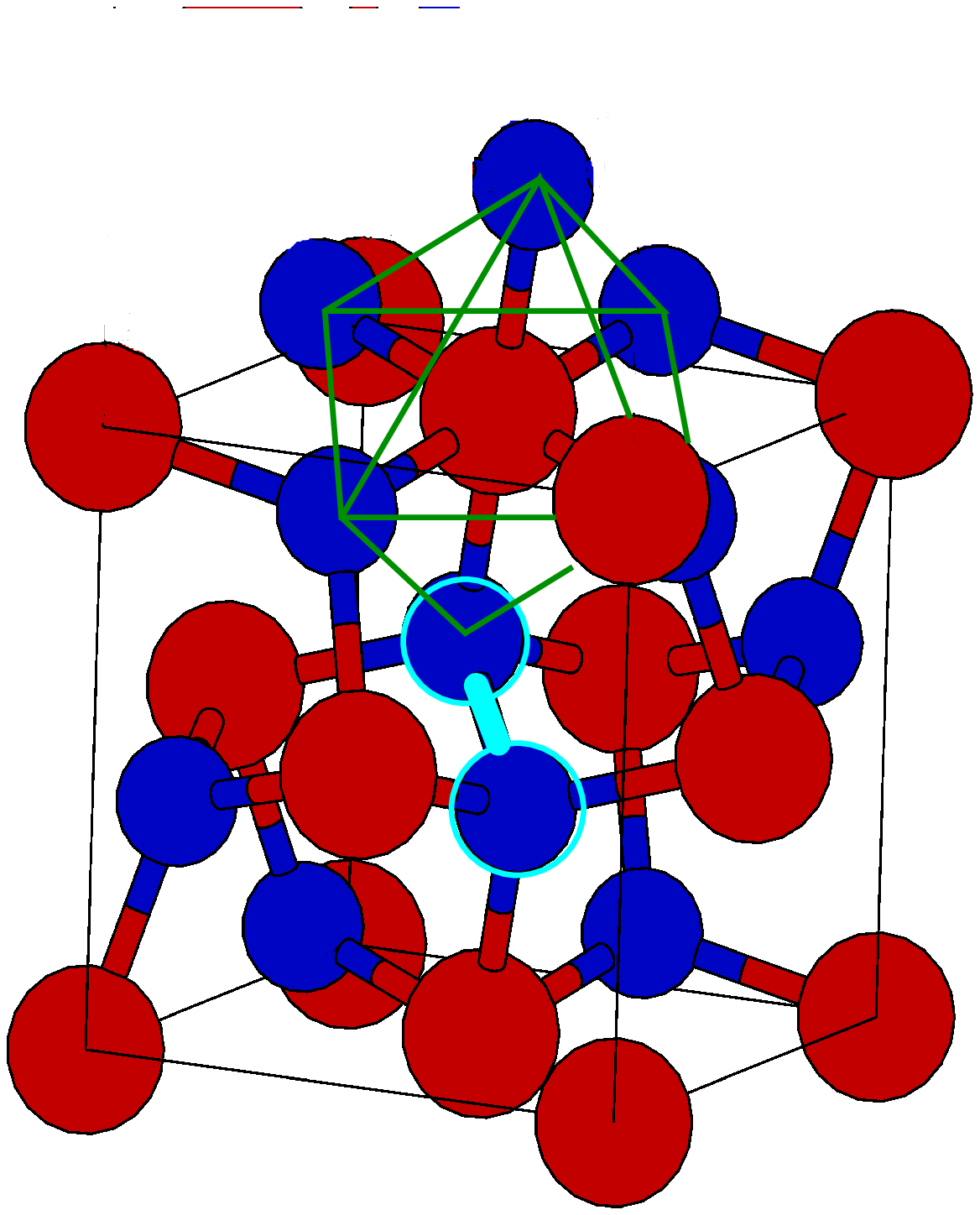}
\caption{\label{fig:4}The crystal structure of NiS$_2$: Ni atoms red (bright), S atoms blue (dark).
The S-S dimer is highlighted in the center of the figure.}
\end{figure}
\begin{figure}[b]
\begin{center}
\includegraphics[scale=.42,angle=270,clip]{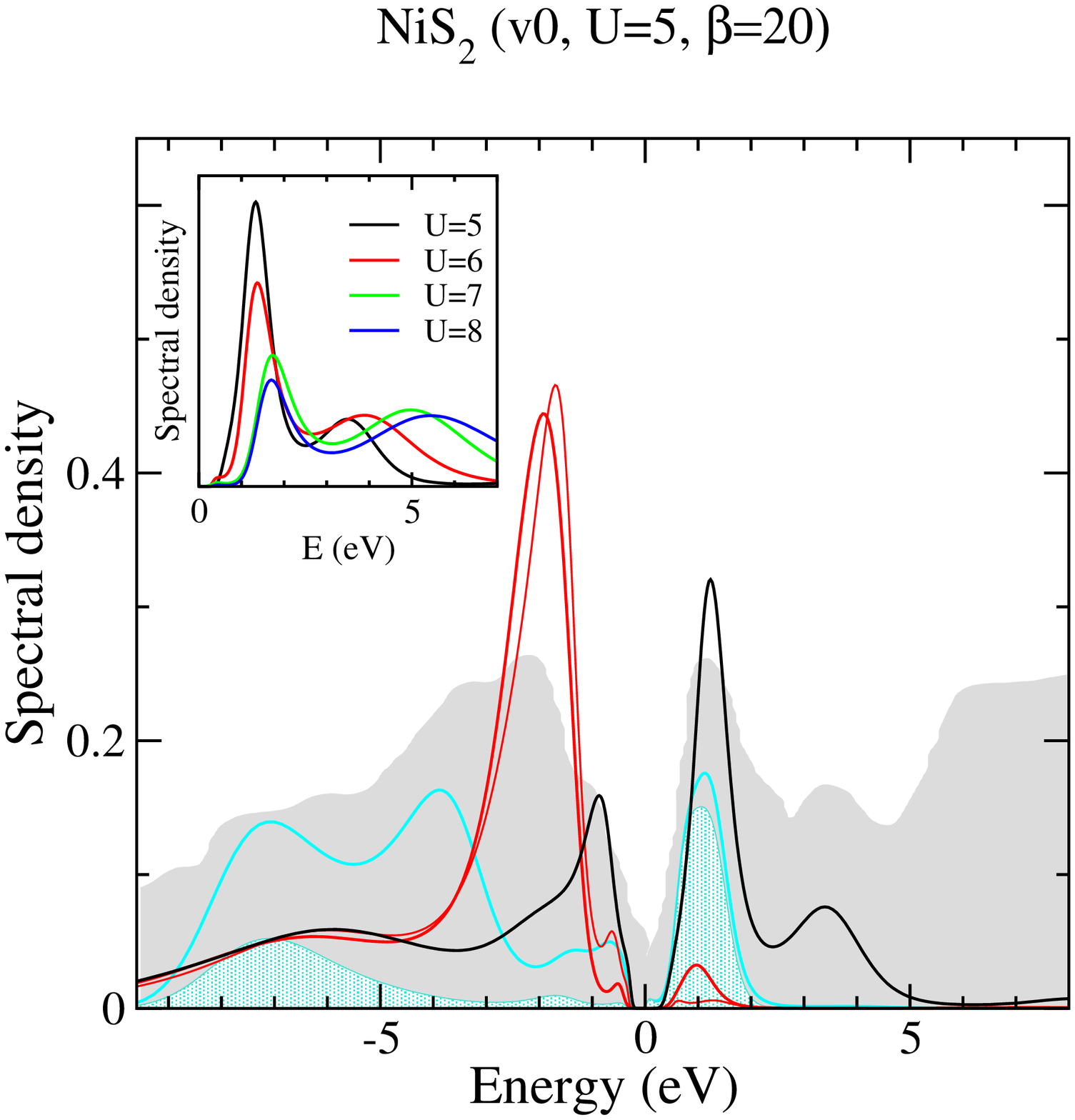}
\end{center}
\caption{\label{fig:5}
The calculated ($U$=5~eV, $T$=580~K) orbitally resolved spectra of NiS$_2$
[Ni-$d$:$e_g^{\sigma}$ (black) and $t_{2g}$=$e_g^{\pi}$ (thick red) + $a_{1g}$ (thin red);
S-$p$: total (blue) and $p_{\sigma}$ (blue-shaded)] compared to
the experimental XPS+BIS data (gray, shaded) \cite{folk87}.
The inset shows the conduction band $e_g^{\sigma}$ spectra for $U$=5, 6, 7, and 8~eV. Figure taken from Ref.\ \cite{kun}.}
\end{figure}
The NiS$_{2-x}$Se$_x$ series presents an important model system in which a metal-insulator transition can be controlled either by varying the 
Se content $x$, temperature $T$, or pressure $P$ \cite{kwi80,yao96,matsu00,miya00,tak07}.
Despite a vast amount of available experimental data a satisfactory material-specific theory for the microscopic
origin of the MIT in NiS$_{2-x}$Se$_x$ is still missing.
Using the LDA+DMFT approach we have discovered an unexpected  
mechanism which controls the opening of the charge gap in NiS$_{2-x}$Se$_x$.

Ni$X$$_2$ ($X$=S, Se) can be viewed as NiO with the O atom replaced by an $X$$_2$ dimer (Fig. \ref{fig:4}), which accommodates
two holes in its $p_{\sigma}^*$ anti-bonding orbitals leading to an $X$$_2^{-2}$ valence state.
It was previously asserted that the empty $p_{\sigma}^*$ bands do not play an active role in the 
physics of Ni$X$$_2$. If true the analogy to NiO becomes complete in the sense that a charge-transfer
gap forms between the ligand $p$-band and the upper Ni-$d$ Hubbard band, whose position is sensitive
to the strength of the on-site Coulomb repulsion $U$. Within such a scenario a stronger screening
leading to smaller $U$ in NiSe$_2$ can explain why NiS$_2$ is an insulator while NiSe$_2$ is a metal.
However, our calculations revealed only a moderate difference in the Coulomb repulsion between NiS$_2$ and NiSe$_2$, and
their respective groundstates turned out to be unaffected by variations of $U$.
This is readily understood by the observation that it is not the upper Ni-$d$ Hubbard band, but rather
the $X$-$p_{\sigma}^*$ band, which forms the bottom of the conduction band (see Fig. \ref{fig:5}).
In the meantime this finding was corroborated by X-ray emission and absorption measurements
at the K-edge of sulfur \cite{kun}.

Our calculations provide the following picture. The different groundstates of NiS$_2$ and NiSe$_2$ are consequence 
of a larger bonding--anti-bonding splitting within the S-S dimer. The splitting in the Se-Se dimer is too small
to open a gap. The closing of the gap in NiS$_2$ by the application of pressure is due to a broadening of the bands while the S-S bond-length related
bonding--anti-bonding splitting remains unchanged since the S-S dimer
behaves as a rigid object in a soft matrix.

\subsection{Interaction Driven Insulator-to-Insulator Transition}

While the Hubbard model has become a paradigm for the description of electronic correlations in metals and the 
metal-insulator transition \cite{Imada98}, much less attention has so far been paid to electronic correlations 
in band insulators. E.g., modeling of Kondo insulators is
far from trivial, and the only recently achieved progress in the topological classification of band insulators 
\cite{Kane05} demonstrates that our understanding of the insulating state is still incomplete.

\begin{figure}[t]
\begin{center}
\vspace{5mm}
\includegraphics[scale=.36,angle=0]{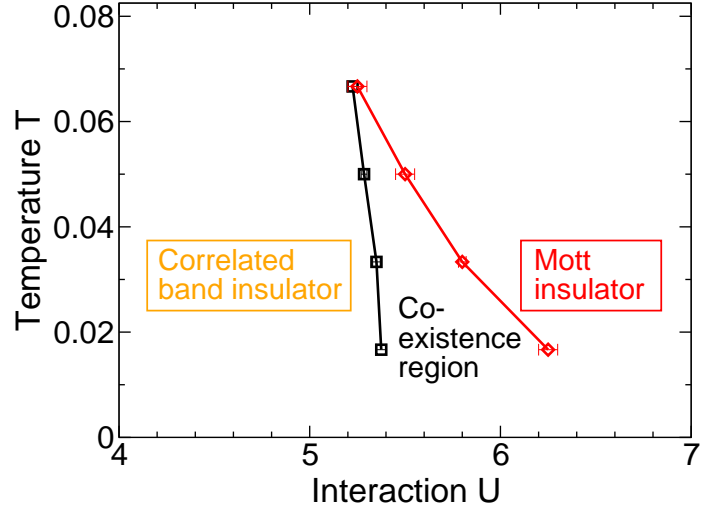}
\end{center}
\caption{The $T$-$U$ phase diagram at fixed $V=0.5$ shows the band and Mott
insulating phases with a first-order phase transition upon increasing the
interaction strength $U$ below a critical temperature. In the coexistence
region both band and Mott insulating solutions of the DMFT equations are
found depending on
the initial guess for the self-energy. For temperatures above the
critical end point of the coexistence region, there is a regime where
the spectral function has a single peak at the Fermi energy accompanied
by broad Hubbard bands. All energies are given in units of $t=1$. Figure taken from \cite{Sentef09}.}
\label{fig:phs}
\end{figure}
Motivated by several investigations of the ionic Hubbard model \cite{Kampf03,Noack04,Aligia04,Garg06,Kancharla07a,Paris07,Craco08,
Byczuk09} we have analyzed a covalent insulator as a complementary
example of a band insulator. As a covalent insulator we denote a band insulator with partially 
filled local orbitals. This definition implies that the 
band gap is a hybridization gap arising from a particular pattern 
of hopping integrals. It has been proposed that similar characteristics apply to
materials such as FeSi, FeSb$_2$ or CoTiSb \cite{Kunes08}, some of which exhibit temperature dependent magnetic and transport properties reminiscent of Kondo insulators. 

In our model study we use a simple particle-hole symmetric model at half-filling described by the Hamiltonian
\beqa
H \!&=&\! \sum_{\BK \sigma} 
\left(
\begin{array}{cc}
a_{\BK\sigma}^{\dagger}, &
b_{\BK\sigma}^{\dagger} \\
\end{array}
\right)
{\bf H}({\BK})
\left(
\begin{array}{cc}
a_{\BK\sigma}^{} \\
b_{\BK\sigma}^{} \\
\end{array}
\right)
+\;U\sum_{i\alpha}n_{i\uparrow\alpha}n_{i\downarrow\alpha} ,
\\
{\bf H}({\BK}) &=& 
\left(
\begin{array}{cc}
\epsilon_{\BK} & V \\
V & - \epsilon_{\BK} \\
\end{array}
\right),
\label{eq:hamilt}
\eeqa
with two semi-circular electronic bands of widths $4t$ ($t=1$ in the following) and dispersions $\epsilon_k$ and $-\epsilon_k$, respectively, corresponding to two sublattices coupled by the $\bm{k}$-independent hybridization $V$ and a local 
electron-electron interaction of strength $U$. Here $n_{i\sigma\alpha}=\alpha_{i\sigma}^{\dagger}\alpha_{i\sigma}^{}$ 
measures the number of electrons with spin $\sigma=\uparrow,\downarrow$ on site $i$ of sublattice $\alpha=a,b$.

\begin{figure}
\begin{center}
\vspace{5mm}
\includegraphics[scale=.36,angle=0]{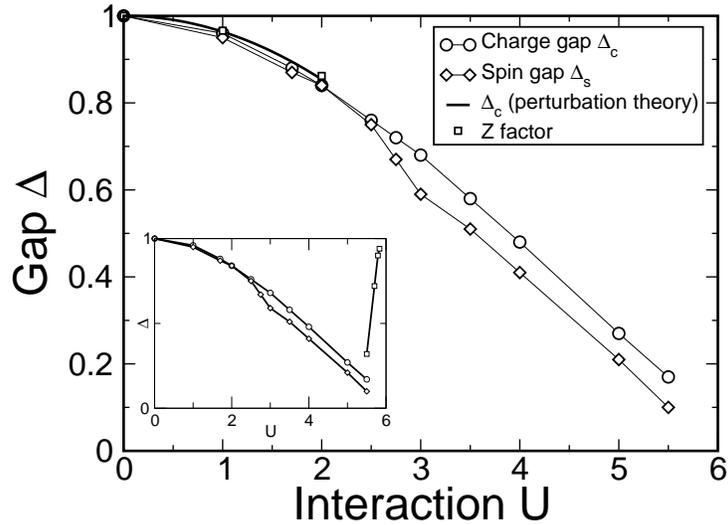}
\end{center}
\caption{
Spin and charge gaps in the correlated band insulator as a function of $U$
for $V=0.5$ as determined from the spectral function and the spin
susceptibility at $T=1/30$, respectively. The thick solid line
shows the charge gap obtained from second order perturbation
theory. The squares represent the $Z$ factor.
Inset: Discontinuous change of spin and charge gaps at the band insulator to Mott insulator transition. In the Mott insulator the spin gap $\Delta_s=0$. All energies are given in units of $t=1$. Figure taken from \cite{Sentef09}.
\label{fig:gap}}
\end{figure}
We use DMFT in conjunction with the recently
developed continuous-time QMC \cite{Werner06} to study the evolution from the band insulator at small to the Mott insulator at large Coulomb interaction strength. The
insulator-insulator transition is discontinuous at finite but low temperatures, with
a region of interaction strengths where hysteretic behavior with two solutions of
the DMFT equations is observed (see Fig.\ \ref{fig:phs}). The behavior of charge and
spin gaps upon increasing the interaction strength is shown in Fig.\ \ref{fig:gap} \cite{Sentef09}. Surprisingly we find that both gaps shrink with increasing Coulomb repulsion. This behavior is in contrast to the correlation-induced Mott
insulator where the charge gap increases with increasing interaction strength.
Furthermore, in the correlated insulator charge and spin gaps deviate from each
other, and the spin gap is smaller than the charge gap. From the self-energy we
extract a renormalization factor $Z$ defined in the same way as the quasiparticle
weight in a Fermi liquid. In the band insulator the $Z$ factor describes the
renormalization of the charge gap at moderate interaction strengths (see Fig.\ \ref{fig:gap}). Thus we obtain
the remarkable finding that a concept from Fermi liquid theory can be applied to
quantify correlation effects in a band insulator!

\section{Conclusions}
We have presented several examples for the application of dynamical
mean-field theory to investigate quantitatively the properties of selected electronically correlated materials. As the main numerical tool a continuous-time quantum Monte-Carlo algorithm was used to solve the
auxiliary impurity problem in the DMFT self-consistency cycle. The present method proved to be a powerful tool to calculate spectral properties of materials with strong electronic correlations, as well as phase transitions near and above room temperature and at arbitrary pressure. The presented model analysis for a covalent band insulator with local Hubbard-type interaction provided new insights into the previously unexplored effects of correlations in band insulators. 

\begin{acknowledgement}
This work was conducted in project pr28je ``Dynamical Mean-Field Theory
for Electronically Correlated Materials'' on the supercomputer HLRB II at the Leibniz-Rechenzentrum in Munich. We thank Philipp Werner for sharing his expertise regarding the continuous-time QMC algorithm. M.S.\
acknowledges support by the Studienstiftung des Deutschen Volkes. The
research was performed within the Sonderforschungsbereich 484 funded by
the Deutsch Forschungsgemeinschaft. The calculations made use of the ALPS library \cite{alps,alps2}. 
\end{acknowledgement}

\bibliographystyle{spphys}
\bibliography{HLRB_Report_Final_Bibliography}

\begin{thebibliography}{10}
\providecommand{\url}[1]{{#1}}
\providecommand{\urlprefix}{URL }
\expandafter\ifx\csname urlstyle\endcsname\relax
  \providecommand{\doi}[1]{DOI \discretionary{}{}{}#1}\else
  \providecommand{\doi}{DOI \discretionary{}{}{}\begingroup
  \urlstyle{rm}\Url}\fi

\bibitem{pt}
G.~Kotliar, D.~Vollhardt, Phys. Today \textbf{57(3)}, 53 (2004)

\bibitem{Imada98}
M.~Imada, A.~Fujimori, Y.~Tokura, Rev. Mod. Phys. \textbf{70}, 1039 (1998)

\bibitem{Georges96}
A.~Georges, G.~Kotliar, W.~Krauth, M.J. Rozenberg, Rev. Mod. Phys. \textbf{68},
  13 (1996)

\bibitem{Metzner89a}
W.~Metzner, D.~Vollhardt, Phys. Rev. Lett. \textbf{62}, 324 (1989)

\bibitem{Georges92a}
A.~Georges, G.~Kotliar, Phys. Rev. B \textbf{45}, 6479 (1992)

\bibitem{Jarrell92}
M.~Jarrell, Phys. Rev. Lett. \textbf{69}, 3410 (1992)

\bibitem{ldadmft}
K.~Held, I.A. Nekrasov, G.~Keller, V.~Eyert, N.~Bl\"umer, A.K. McMahan, R.T.
  Scalettar, T.~Pruschke, V.I. Anisimov, D.~Vollhardt, Phys. Status Solidi B
  \textbf{243}, 2599 (2006)

\bibitem{Werner06}
P.~Werner, A.~Comanac, L.~de' Medici, M.~Troyer, A.J. Millis, Phys. Rev. Lett.
  \textbf{97}, 076405 (2006)

\bibitem{Gull07}
E.~Gull, P.~Werner, A.J. Millis, M.~Troyer, Phys. Rev. B \textbf{76}, 235123
  (2007)

\bibitem{Alvarez08}
G.~Alvarez, M.S. Summers, D.E. Maxwell, M.~Eisenbach, J.S. Meredith, J.M.
  Larkin, J.~Levesque, T.A. Maier, P.R.C. Kent, E.F. D'Azevedo, T.C.
  Schulthess, Article No. 61 in \emph{Proceedings of the 2008 ACM/IEEE
  conference on Supercomputing} pp. 1--10 (2008)

\bibitem{alps}
M.~Troyer, B.~Ammon, E.~Heeb, Lect. Notes Comput. Sci. \textbf{1505}, 191
  (1998)

\bibitem{lad-89}
R.J. Lad, V.E. Henrich, Phys. Rev. B \textbf{39}, 13478 (1989)

\bibitem{ciccacci-91}
F.~Ciccacci, L.~Braicovich, E.~Puppin, E.~Vescovo, Phys. Rev. B \textbf{44},
  10444 (1991)

\bibitem{fe2o3}
J.~Kune\v{s}, D.M. Korotin, M.A. Korotin, V.I. Anisimov, P.~Werner, Phys. Rev.
  Lett. \textbf{102}, 146402 (2009)

\bibitem{yoo}
C.S. Yoo, B.~Maddox, J.H.P. Klepeis, V.~Iota, W.~Evans, A.~McMahan, M.Y. Hu,
  P.~Chow, M.~Somayazulu, D.~H\"ausermann, W.E. Pickett, R.T. Scalettar, Phys.
  Rev. Lett. \textbf{94}, 115502 (2005)

\bibitem{gav-08}
A.G. Gavriliuk, V.V. Struzhkin, I.S. Lyubutin, S.G. Ovchinnikov, M.Y. Hu,
  P.~Chow, Phys. Rev. B \textbf{77}, 155112 (2008)

\bibitem{Sgull-51}
C.G. Shull, W.A. Strauser, E.O. Wollan, Phys. Rev. \textbf{83}, 333 (1951)

\bibitem{Fujimori-86}
A.~Fujimori, M.~Saeki, N.~Kimizuka, M.~Taniguchi, S.~Suga, Phys. Rev. B
  \textbf{34}, 7318 (1986)

\bibitem{mochizuki}
S.~Mochizuki, Phys. Status Solidi A \textbf{41}, 591 (1977)

\bibitem{Pasternak-99}
M.~Pasternak, G.~Rozenberg, G.~Machavariani, O.~Naaman, R.~Taylor, R.~Jeanloz,
  Phys. Rev. Lett. \textbf{82}, 4663 (1999)

\bibitem{Rozenberg-02}
G.~Rozenberg, L.~Dubrovinsky, M.~Pasternak, O.~Naaman, T.L. Bihan, R.~Ahuja,
  Phys. Rev. B \textbf{65}, 064112 (2002)

\bibitem{Liu-03}
H.~Liu, W.A. Caldwell, L.R. Benedetti, W.~Panero, R.~Jeanloz, Phys. Chem.
  Miner. \textbf{30}, 582 (2003)

\bibitem{Badro-02}
J.~Badro, G.~Fiquet, V.~Struzhkin, M.~Somayazulu, H.K. Mao, G.~Shen, T.L.
  Bihan, Phys. Rev. Lett. \textbf{89}, 205504 (2002)

\bibitem{yb}
E.R. Ylvisaker, J.~Kune\v{s}, A.K. McMahan, W.E. Pickett, Phys. Rev. Lett.
  \textbf{102}, 246401 (2009)

\bibitem{strange_nature}
P.~Strange, A.~Svane, W.M. Temmerman, Z.~Szotek, H.~Winter, Nature
  \textbf{399}, 756 (1999)

\bibitem{Takemura85}
K.~Takemura, K.~Syassen, J. Phys. F \textbf{15}, 543 (1985)

\bibitem{Barson57}
F.~Barson, S.~Legvold, F.H. Spedding, Phys. Rev. \textbf{105}, 418 (1957)

\bibitem{folk87}
W.~Folkerts, G.A. Sawatzky, C.~Haas, R.A. de~Groot, F.U. Hillebrecht, J. Phys.
  C \textbf{20}, 4135 (1987)

\bibitem{kun}
J.~{Kune\v{s} {\em et al.}}, To be published

\bibitem{kwi80}
P.~Kwizera, M.S. Dresselhaus, D.~Adler, Phys. Rev. B \textbf{21}, 2328 (1980)

\bibitem{yao96}
X.~Yao, J.M. Honig, T.~Hogan, C.~Kannewurf, J.~Spa{\l}ek, Phys. Rev. B
  \textbf{54}, 17469 (1996)

\bibitem{matsu00}
M.~Matsuura, H.~Hiraka, K.~Yamada, Y.~Endoh, J. Phys. Soc. Jpn. \textbf{69},
  1503 (2000)

\bibitem{miya00}
S.~Miyasaka, H.~Takagi, Y.~Sekine, H.~Takahashi, N.~Mori, R.J. Cava, J. Phys.
  Soc. Jpn. \textbf{69}, 3166 (2000)

\bibitem{tak07}
N.~Takeshita, S.~Takashima, C.~Terakura, H.~Nishikubo, S.~Miyasaka, M.~Nohara,
  Y.~Tokura, H.~Takagi.
\newblock ArXiv:0704.0591

\bibitem{Kane05}
C.L. Kane, E.J. Mele, Phys. Rev. Lett. \textbf{95}, 146802 (2005)

\bibitem{Sentef09}
M.~Sentef, J.~Kune\v{s}, P.~Werner, A.P. Kampf, Phys. Rev. B \textbf{80},
  155116 (2009)

\bibitem{Kampf03}
A.P. Kampf, M.~Sekania, G.I. Japaridze, P.~Brune, J. Phys. Condens. Matter
  \textbf{15}, 5895 (2003)

\bibitem{Noack04}
S.R. Manmana, V.~Meden, R.M. Noack, K.~Sch\"onhammer, Phys. Rev. B \textbf{70},
  155115 (2004)

\bibitem{Aligia04}
C.D. Batista, A.A. Aligia, Phys. Rev. Lett. \textbf{92}, 246405 (2004)

\bibitem{Garg06}
A.~Garg, H.R. Krishnamurthy, M.~Randeria, Phys. Rev. Lett. \textbf{97}, 046403
  (2006)

\bibitem{Kancharla07a}
S.S. Kancharla, E.~Dagotto, Phys. Rev. Lett. \textbf{98}, 016402 (2007)

\bibitem{Paris07}
N.~Paris, K.~Bouadim, F.~Hebert, G.G. Batrouni, R.T. Scalettar, Phys. Rev.
  Lett. \textbf{98}, 046403 (2007)

\bibitem{Craco08}
L.~Craco, P.~Lombardo, R.~Hayn, G.I. Japaridze, E.~M\"uller-Hartmann, Phys.
  Rev. B \textbf{78}, 075121 (2008)

\bibitem{Byczuk09}
K.~Byczuk, M.~Sekania, W.~Hofstetter, A.P. Kampf, Phys. Rev. B \textbf{79},
  121103(R) (2009)

\bibitem{Kunes08}
J.~Kune\v{s}, V.I. Anisimov, Phys. Rev. B \textbf{78}, 033109 (2008)

\bibitem{alps2}
F.~Alet, P.~Dayal, A.~Grzesik, A.~Honecker, M.~K\"{o}rner, A.~L\"{a}uchli, S.R.
  Manmana, I.P. McCulloch, F.~Michel, R.M. Noack, G.~Schmid, U.~Schollw\"{o}ck,
  F.~St\"{o}ckli, S.~Todo, S.~Trebst, M.~Troyer, P.~Werner, S.~Wessel, J. Phys.
  Soc. Jpn. Suppl. \textbf{74}, 30 (2005)

\end{thebibliography}
\end{document}